\documentclass[reprint, nofootinbib, aps, preprintnumbers ]{revtex4-2}
\usepackage{amsfonts}
\usepackage{amsmath,amssymb}
\usepackage{bbm}
\usepackage{bm}
\usepackage{color}
\usepackage{slashed}
\usepackage[utf8]{inputenc}
\usepackage[toc,page]{appendix}
\usepackage{graphicx}
\usepackage{placeins}
\usepackage{caption}
\usepackage{subfigure}
\usepackage{booktabs}
\usepackage{comment}
\usepackage{physics}
\usepackage{float}
\usepackage{bbold}
\usepackage{soul}
\usepackage{bm}
\usepackage{ragged2e}
\usepackage{dcolumn}
\DeclareGraphicsExtensions{.pdf,.png,.jpg}

\usepackage[colorlinks]{hyperref}
\AtBeginDocument{%
  \hypersetup{
    citecolor=blue,
    linkcolor=blue,   
    urlcolor=blue}}

\definecolor{airforceblue}{rgb}{0.36, 0.54, 0.66}
\definecolor{steelblue}{rgb}{0.27, 0.51, 0.71}
\definecolor{amber}{rgb}{1.0, 0.49, 0.0}
\definecolor{darkgreen}{rgb}{0.0, 0.5, 0.0}
\definecolor{amber}{rgb}{1.0, 0.49, 0.0}

\DeclareMathAlphabet{\mathpzc}{OT1}{pzc}{m}{it}

\graphicspath{{draft/Figures/}}

\def\simg{{\ \lower-1.2pt\vbox{\hbox{\rlap{$>$}\lower6pt\vbox{\hbox{$\sim$}}}}\ }}
\def\siml{{\ \lower-1.2pt\vbox{\hbox{\rlap{$<$}\lower6pt\vbox{\hbox{$\sim$}}}}\ }}

\makeatletter \@addtoreset{equation}{section} \makeatother

\graphicspath{{Figures/}}

\begin{document}
\preprint{IFT-UAM/CSIC-22-131}

\title{
A lesson from $R_{\tau\tau}^{K^{(\ast)}}$ and $R_{\nu\nu}^{K^{(\ast)}}$ at Belle~II}

\author{Arturo de Giorgi}\email{arturo.degiorgi@uam.es}
\affiliation{%
 Departamento de F\'isica Te\'orica and Instituto de F\'isica Te\'orica UAM/CSIC,\\
Universidad Aut\'onoma de Madrid, Cantoblanco, 28049, Madrid, Spain
}%

\author{Gioacchino Piazza}\email{gioacchino.piazza@ijclab.in2p3.fr}\affiliation{ IJCLab, P\^ole Th\'eorie (Bat.~210), CNRS/IN2P3 et Universit\'e, Paris-Saclay, 91405 Orsay, France}
   \begin{abstract}
   Within the assumption of Left-Handed (LH) New Physics (NP), we review the relations between $\mathcal{B}(B\to K^{(\ast)} \tau^+\tau^-)$ and $\mathcal{B}(B\to K^{(\ast)} \nu\bar \nu)$ 
    for several Beyond the Standard Model (BSM) scenarios, 
    commonly considered to explain the Lepton flavor Universality (LFU) violation observed in charged and neutral-current semileptonic $B$ decays. 
  We employ the latest $R_{D^{(\ast)}}$ world averages that include the recent LHCb measurement and assess the possibility of simultaneously explaining the $B$-anomalies without spoiling current bounds on di-neutrino and di-tau modes.
   This is particularly relevant in light of the upcoming results by Belle II on neutrinos and the continuing improvement in accuracy and sensitivity achieved in tau modes.
   
   \end{abstract}
\maketitle
\begin{section}{Introduction}\label{sec:intro}
Flavor Physics has been in the last decades a forerunner in the search for NP. As flavor-changing neutral currents (FCNC) are loop and GIM suppressed in the Standard Model (SM), the related observables are supposed to be very sensitive to contributions from NP.

In recent years, a number of $B$-meson-related anomalies have appeared (see~\cite{Guadagnoli:2022oxk} for a recent review). Among these, two are probably the best known and most transparent in their interpretation. The first one concerns deviations in the ratios 
$R_{K^{(*)}}\equiv \mathcal{B}^\prime(B\to K^{(*)}\mu^+\mu^-)/\mathcal{B}^\prime(B\to K^{(*)}e^+e^-)$ with respect to the SM value, where $\mathcal{B}'$ stands for binned branching ratio. The latest measurements of LHCb have found
\begin{align}
&R_K^{[1.1,6.0]\text{GeV}^2}=0.846^{+0.042}_{-0.039}{}^{+0.013}_{-0.012} & \text{\cite{LHCb:2021trn}}
\label{ExperimentalRK}\\
&R_{K^\ast}^{[1.1,6.0]\text{GeV}^2}=
0.69^{+0.11}_{-0.07}{}\pm0.05  & \text{\cite{LHCb:2017avl}}
\\
&R_{K^\ast}^{[0.045,1.1]\text{GeV}^2}= 0.66^{+0.11}_{-0.07}{}\pm0.03  & \text{\cite{LHCb:2017avl}}
\label{ExperimentalRKstar}
\end{align}
The SM predictions read $1.00\pm0.01$ for $R_{K^{(*)}}$ in the bin $q^2 \in [1.1,6.0]$ GeV$^2$, and $0.92\pm 0.02$ in the bin $[0.045,1.1]$ GeV$^2$~\cite{Hiller:2003js,Bobeth:2007dw,Bordone:2016gaq, Isidori:2020acz}, showing several discrepancies between $\sim 2\sigma$ and $\sim 3\sigma$. 

The second anomalies are related to charged currents and show deviations in the ratios $R_{D^{(*)}}\equiv \mathcal{B}(B\to D^{(*)}\tau^-\bar{\nu}_\tau)/\mathcal{B}(B\to D^{(*)}\ell^-\bar{\nu}_\ell)$, with $\ell=e,\mu$.
Recently, an updated measurement of $R_{D}$ and $R_{D^{*}}$ has been announced by LHCb~\cite{LHcb-partial}, which shows agreement with the previous measurement~\cite{Amhis:2022mac}. The preliminary HFLAV 2022 averages read
\begin{align}
    &R_D = 0.358 \pm 0.025 \pm 0.012 \,,\\
    &R_{D^*} = 0.285 \pm 0.010 \pm 0.008 \,,
\end{align}
which have to be compared to the SM predictions~\cite{LHcb-partial}
\begin{align}
    &R_D^\text{SM} = 0.298 \pm 0.004 \,,\\
    &R_{D^*}^\text{SM} = 0.254 \pm 0.005 \,,
\end{align}
showing an overall tension at the $3.2 \, \sigma$ level (see \cite{Iguro:2022yzr} for a recent phenomenological analysis). 

Several BSM scenarios have been discussed in the literature~\cite{Bauer:2015knc, Fajfer:2015ycq, Barbieri:2015yvd, Greljo:2015mma, BHATTACHARYA2015370,  Becirevic:2016yqi,  Boucenna:2016qad, Alok:2017jaf, Crivellin:2017zlb, Assad:2017iib, Buttazzo:2017ixm, Bordone:2017bld,DiLuzio:2017chi, DiLuzio:2017vat,Dorsner:2017ufx,Becirevic:2017jtw, Becirevic:2018afm,DiLuzio:2018zxy,Bordone:2018nbg,  Matsuzaki:2018jui, Crivellin:2018yvo, Angelescu:2018tyl, Blanke:2018sro, Kumar:2018kmr, Cornella:2019hct, Popov:2019tyc, Bigaran:2019bqv, Hati:2019ufv, Altmannshofer:2020axr,  Angelescu:2021lln,Bonilla:2022qgm}, and it is quite challenging to formulate a scenario that would be consistent with both $R_{D^{(\ast)}}$ and $R_{K^{(\ast)}}$ and with a wealth of low energy flavor physics observables.

The global fits of the neutral $B$-anomalies seem to favor NP scenarios coupling predominantly to the left-handed muons, and not affecting the electrons~\cite{Ciuchini:2019usw, Kowalska:2019ley, Datta:2019zca,Alguero:2021anc,Aebischer:2017gaw}. Motivated by this hint, it is then a reasonable assumption to consider that NP may only couple to left-handed fermions. This would somewhat be a replica of what happens with the $SU(2)_L$ gauge-symmetry of the SM, making it an interesting possibility from the theoretical point of view. 

The anomalies in $R_{D^{(*)}}$ seem to suggest modifications also in the $\tau$ sector. Since $b\to c \tau \nu_\tau$ is a tree-level process in the SM, the NP contribution must be sizeable and much bigger than what is required to explain $R_{K^{(\ast)}}$.
In fact, processes involving taus have already been investigated in depth in the literature~\cite{Alonso:2015sja,Feruglio:2016gvd, Capdevila:2017iqn,Crivellin:2017zlb,Calibbi:2017qbu, Cornella:2018tfd, Cornella:2021sby,Becirevic:2012jf}.
As the experimental sensitivity to $\tau$ is considerably smaller than the one for electrons and muons, the possibility of large contributions to
\begin{equation}
    R^{K^{(\ast)}}_{\tau\tau} \equiv \frac{\mathcal{B}(B\to K^{(\ast)} \tau\tau)}{ \mathcal{B}(B\to K^{(\ast)} \tau\tau)_{\rm SM}}\,
\end{equation}
remains viable.
The values expected for such observable in different models are typically very large, even of $\sim\mathcal{O}(700)$~\cite{Capdevila:2017iqn}, as they are dominated by the large NP contribution required by $R_{D^{(*)}}$.

On the other hand, the same operators affecting $R_{D^{(*)}}$ usually generate a large impact on $b \to s \nu\nu$, unless some cancellations between the Wilson coefficients happen.
The most constraining bounds on di-neutrino modes come from the Belle collaboration, namely~\cite{Belle:2017oht} 
\begin{align}
  &\label{eq:belleBounds1}
  R^{K}_{\nu\nu} < 3.9 \quad(90\% \ \text{C.L.})\,,\\
 &\label{eq:belleBounds2}
 R^{K^*}_{\nu\nu} < 2.7 \quad(90\% \ \text{C.L.})\,,
\end{align}
where $R^{K^{(\ast)}}_{\nu\nu}$ indicates the ratio between the upper bounds on $\mathcal{B}(B\to K^{(\ast)} \nu\nu)$ and the respective SM prediction, combining charged and
neutral modes.

In this letter, we focus on the relations between $b \to s \nu\nu$ and $b \to s \tau\tau$ generated in various models assuming that
\begin{itemize}
    \item NP enters only through LH operators,
    \item NP couples diagonally to leptons,
    \item NP couples negligibly to electrons.
\end{itemize}
We re-examine the viability of many scenarios involving one BSM field, leptoquarks or Vector-Boson (VB), in the light of the most recent $R_{D^{(*)}}$ measurement~\cite{LHcb-partial}, and the constraints from $R^{K^{(\ast)}}_{\nu\nu} $.
We furthermore identify possible two-field extensions that can explain the $B$-anomalies, and at the same time generate $R^{K^{(\ast)}}_{\tau\tau}$ and $R^{K^{(\ast)}}_{\nu\nu}$ of $\mathcal{O}(1)$.
This is particularly interesting in light of the prospects of Belle II~\cite{Belle-II:2018jsg, Belle-II:2022cgf}, which within the current decade is expected to reach an integrated luminosity of $50 \ \text{ab}^{-1}$.
According to the Refs.~\cite{Belle-II:2018jsg, Belle-II:2022cgf}, Belle II should reach a sensitivity with respect to the SM predictions of $0.55(0.11)$ on $B^+\to K^+\nu\bar{\nu}$ and $1.08(0.34)$ on $B^0\to K^{0*}\nu\bar{\nu}$, for an integrated luminosity of $1\ \text{ab}^{-1}(50\ \text{ab}^{-1})$.\footnote{The current combined integrated luminosity of the data from Belle and Belle II is about $1\ \text{ab}^{-1}$\cite{Belle-II:2022cgf}.} The sensitivities for $B\to K^{(\ast)} \tau \tau$ are also expected to improve roughly by a factor of three. In the most optimistic case, Belle II could start probing $R_{\tau\tau}^{K^\ast}\sim\mathcal{O}(4000)$ with $50\ \text{ab}^{-1}$ of integrated luminosity \cite{Belle-II:2022cgf}, which is however far from the SM value and leaves space for NP models with large taus contributions.

The paper is structured as follows. In Sec.~\ref{Sec:SMEFT} we introduce the necessary theoretical framework, with a particular focus on the LH operators needed to explain the anomalies. In Sec.~\ref{sec:Models} we introduce the relevant single-field extensions and their connection to the operators of interest. Finally in Sec.~\ref{sec:Analysis} we carry out the analysis and draw conclusions in Sec.~\ref{sec:Conlusions}.
\end{section}
\begin{section}{SMEFT and LEFT}\label{Sec:SMEFT}
If the scale of NP is above the Electroweak scale, the SM Effective Field Theory (SMEFT) is a powerful tool to examine in a model-independent way NP contributions to low-energy observables.
 The SMEFT provides a framework in which the operators invariant under  $SU(3)_c\times  SU(2)_L \times U(1)_Y$ are organized according to their mass dimension~\cite{smeft1,BUCHMULLER1986621, Grzadkowski:2010es}. 

The Lagrangian containing the dimension 6 operators reads
\begin{equation}
  \mathcal{L}^{(6)}=\sum_i \frac{C_i}{\Lambda^2} \mathcal{O}_i  \,,
\end{equation}
where $\Lambda$ stands for the NP scale.
Consistently with the hypothesis of LH NP only, the 4-fermions operators relevant to our study are\footnote{The other 4-fermions operators include right-handed fermions and are thus not considered here. 
} 
\begin{equation}
\begin{split}
\big{[}\mathcal{O}_{lq}^{(1)}\big{]}_{ijkl} &= \big{(}\overline{L}_i\gamma^\mu L_j\big{)} \big{(}\overline{Q}_k \gamma_\mu Q_l\big{)}\,, \\[0.4em]
\big{[}\mathcal{O}_{lq}^{(3)}\big{]}_{ijkl} &=\big{(}\overline{L}_i\gamma^\mu \tau^ I L_j\big{)} \big{(}\overline{Q}_k \tau^ I\gamma_\mu Q_l\big{)}\,,
\end{split}
\end{equation}
where $\lbrace Q, L\rbrace$ denote the SM quark and lepton $SU(2)_L$ doublets. The Pauli matrices $\tau^I$ act on the weak indices, while flavor indices are denoted by $\lbrace i,j,k,l \rbrace$.  We adopt the basis defined by the diagonal down-quark Yukawa matrix, and the quark doublet given by 
\begin{equation}
\label{eq:change-basis}
    Q_i=[(V_{\rm CKM}^\dagger\,u)_i\, ,\,d_i ]^T\,.
\end{equation}

After the Electroweak symmetry breaking, these operators contribute to semileptonic $B$ decays, which are well described by the Low-Energy Effective Field Theory (LEFT). 
For each family, we write the total LEFT coefficient as
\begin{equation}
    C_i = C_i^\text{SM}+ \delta C_i \,,
\end{equation}
where $ C_i^\text{SM}$ is the value of the Wilson coefficient generated in the SM, and $\delta  C_i$ is the NP contribution.
The transitions that we will consider in this study are $b\to s \ell \ell$, $b\to s \nu \nu$ and $b\to c \tau \nu_\tau$.

\subsection*{Matching}\label{sec:treelevelmatching}
\paragraph*{\underline{\bm{$b\to s\ell \ell$}}}
The effective Lagrangian describing the $b\to s \ell \ell$ transition is 
\begin{align}
\label{eq:Lbsll}
\mathcal{L}_\mathrm{eff}^{b\to s \ell \ell} =  \dfrac{4 G_F}{\sqrt{2}} \lambda_t \frac{\alpha_{\rm em}}{4\pi}\sum_a C_a\, \mathcal{O}_a+\mathrm{h.c.}\,, 
\end{align}

\noindent where $G_F$ is the Fermi constant, $|\lambda_t| \equiv |V_{tb} V_{ts}^\ast|= 0.040(1)$~\cite{Aoki:2021kgd} is the product of CKM matrix elements,  
and the relevant operators are
  
\begin{align}
\mathcal{O}_{9 }^{(\prime)b s\ell \ell} &=  \big{(}\bar{s}\gamma_\mu P_{L(R)} b \big{)}\big{(}\bar{\ell}\gamma_\mu \ell \big{)}\,,\label{eq:O9}\\
\mathcal{O}_{10}^{(\prime) bs\ell \ell} &=  \big{(}\bar{s}\gamma_\mu P_{L(R)} b \big{)}\big{(}\bar{\ell}\gamma_\mu\gamma_5 \ell \big{)}\,\label{eq:O10}\,,
\end{align}
in addition to the dipole operators $\mathcal{O}_{7,8}$~\cite{Bobeth:1999mk,Altmannshofer:2008dz}. 
The SM Wilson coefficients can be found in Ref.~\cite{Altmannshofer:2008dz}

\noindent
The tree-level matching to the SMEFT gives~\cite{Alonso:2014csa}
\begin{equation}
    \begin{split}
\label{eq:bsll-smeft}
    &\delta C_9^{\ell \ell}=-\delta C_{10}^{\ell \ell} = \dfrac{\pi}{\alpha_\mathrm{em} \lambda_t} \dfrac{v^2}{\Lambda^2} \left\lbrace\big{[}C_{lq}^{(1)}\big{]}_{\ell \ell 23}+\big{[}C_{lq}^{(3)}\big{]}_{\ell \ell 23}\right\rbrace\,, \\[0.4em]
    &\delta C'{}_9^{\ell \ell}=\delta C'{}_{10}^{\ell \ell} =0\,.
    \end{split}
\end{equation}
The condition $\delta C_{9}^{\ell \ell}=-\delta C_{10}^{\ell \ell}$ can be violated only in presence of the Right-Handed (RH) NP. Similarly, $\delta C^{\prime}_{9,10}\neq 0$ only if RH NP is present.
\paragraph*{\underline{\bm{$b\to s\nu \nu$}}}
The $b\to s \nu\nu$ transition can be described by the Lagrangian in Eq.~(\ref{eq:Lbsll}), where this time the relevant operators are given by \cite{Buras:2014fpa}
\begin{align}
\mathcal{O}_{L}^{\nu_i\nu_j} &=(\bar{s}_L \gamma_\mu b_L)(\bar{\nu}_i \gamma^\mu (1-\gamma_5)\nu_j)\,, \\
\mathcal{O}_{R}^{\nu_i\nu_j} &=(\bar{s}_R \gamma_\mu b_R)(\bar{\nu}_i \gamma^\mu (1-\gamma_5)\nu_j).
\end{align}

\noindent The tree-level matching to the SMEFT gives
\begin{align}
\begin{split}
\label{eq:left-CL-bsnunu}
\delta C_L^{\nu_i \nu_j} &= \dfrac{\pi}{\alpha_\mathrm{em} \lambda_t} \dfrac{v^2}{\Lambda^2} \left\lbrace\big{[}C_{lq}^{(1)}\big{]}_{ij23}-\big{[}C_{lq}^{(3)}\big{]}_{ij23}\right\rbrace \,, \\[0.4em]
\delta C_R^{\nu_i \nu_j} &=0\,.
\end{split}
\end{align}
Again, $\delta C_R^{\nu_i \nu_j} \neq0$ only if RH physics is included.

\paragraph*{\underline{\bm{$b\to c \tau \nu_\tau$}}}
The effective Lagrangian for the 
$b\to c \tau \nu_\tau$ process is~\cite{Bardhan:2016uhr}
\begin{align}
\label{eq:Lbctv}
\mathcal{L}_\mathrm{eff}^{b\to c \tau \nu_\tau} =-  \dfrac{4 G_F}{\sqrt{2}} V_{cb} \sum_a C_a\, \mathcal{O}_a+\mathrm{h.c.}\,, 
\end{align}
and the relevant operators are 
\begin{align}
\mathcal{O}_V^{bc\tau \nu} &=  \big{(}\bar{c}\gamma_\mu P_{L} b \big{)}\big{(}\bar{\tau}\gamma_\mu \nu \big{)}\,,\label{eq:O9nu}\\
\mathcal{O}_{A}^{bc\tau \nu} &=  \big{(}\bar{c}\gamma_\mu P_{L} b \big{)}\big{(}\bar{\tau}\gamma_\mu\gamma_5 \nu \big{)}\,\label{eq:O10nu}\,.
\end{align}
The tree-level matching to the SMEFT Lagrangian finally reads
\begin{equation}
    \begin{split}
\label{eq:bslnu-smeft}
    \delta C_V^{\tau \nu}-\delta C_{A}^{\tau \nu} &=\text{$-$}\dfrac{v^2}{V_{cb} \Lambda^2}\big{[}C_{lq}^{(3)}\big{]}_{3323}\,, \\[0.4em]
    \end{split}
\end{equation}
where we neglected $\big{[}C_{lq}^{(3)}\big{]}_{3313}$ and $\big{[}C_{lq}^{(3)}\big{]}_{3333}$, which are constrained by  $B^- \to  \tau^- \bar\nu_\tau$ and the studies of the high $p_T$ tails of $pp$ scattering with $\tau^+ \tau^-$ in the final states~\cite{Faroughy:2016osc, Cornella:2021sby,Greljo:2017vvb, Allwicher:2022mcg,Allwicher:2022gkm}, as well as CKM suppressed.

\section{Models and Benchmarks}
\label{sec:Models}
From the matching of the SMEFT to the LEFT, some operators can impact simultaneously both charged and neutral $b$ transitions. In particular, the operator  $\mathcal{O}_{lq}^{(3)}$  contributes to $b\to s \tau \tau $, $b\to s \nu \nu$ and  $b\to c \tau \nu$, while, at tree-level,
$\mathcal{O}_{lq}^{(1)}$
only contributes to $b\to s \tau \tau $ and $b\to s \nu \nu$. 
From Eq.~\eqref{eq:bslnu-smeft} it is evident that, within the assumptions of LH NP only, in order to solve the $R_{D^{(\ast)}}$ anomalies $\big{[}C_{lq}^{(3)}\big{]}_{\ell\ell23}\neq 0$ is required.
Typically, once a UV model is specified, the Wilson coefficients of the SMEFT
are related in such a way that for the LH operators we can write
\begin{equation}
    \big{[}C_{lq}^{(1)}\big{]}_{\ell\ell23} = \kappa_\ell \big{[}C_{lq}^{(3)}\big{]}_{\ell\ell23}\,,
\end{equation}
with $\kappa_\ell$ depending on the UV model (see e.g. Ref.s~\cite{Buras:2014fpa,deBlas:2017xtg, Aebischer:2020mkv,Husek:2021isa}).

We start by considering the simplest possibility, namely that the NP scenario consists of a single new field. The candidates compatible with the hypothesis of this work are listed in Tab.~\ref{tab:models}.
\begin{table}[t!]
\centering
\begin{tabular}{lcr}
\toprule
Model & $\kappa_\ell$ &  \\
\midrule
$\text{VB}\sim (\mathbf{1},\mathbf{3},0)$ & $0^\ast$&   \\
$U_1\sim (\mathbf{3},\mathbf{1},2/3)$ & $1^\ast$& \\
$U_3\sim (\mathbf{3},\mathbf{3},2/3)$ & $-3$&\\
$S_3\sim (\bar{\mathbf{3}},\mathbf{3},1/3)$ & $3$& \\
\midrule
$Z'\sim (\mathbf{1},\mathbf{1},0)$ & $-$&~\\
$S_1\sim (\bar{\mathbf{3}},\mathbf{1},1/3)$ &  $-1^\ast$ &\\
\bottomrule
\end{tabular}
\caption{\small \sl Benchmark models. The $-$ indicates that no relation exists between the SMEFT Wilson coefficients. The models with a star, $\kappa^\ast$, generate also operators involving right-handed leptons: we do not consider their impact here.}\label{tab:models} 
\end{table}
Nonetheless, some of them can be ruled out. The $Z'$ model is obviously not suitable to explain the $R_{D^{(*)}}$ anomalies, since it predicts $\big{[}C_{lq}^{(3)}\big{]}_{3323}=0$. Similarly, the $S_1$ predicts at tree level $\kappa=-1$, which implies $\delta C_9^{\ell\ell}=\delta C_{10}^{\ell\ell}=0$, if no right-handed operator is involved. 

Beyond the extensions involving a single BSM field, combining two-field allows in principle to generate any value of $\kappa$. 
As a proof of concept that such a statement is true, it is enough to consider an extension that includes the $Z'$ and a VB, so that the generation of $\big{[}C_{lq}^{(1,3)}\big{]}_{3323}$ is independent of one another and any value of $\kappa$ can be obtained by choosing the appropriate couplings. Models with combinations of multiple fields not dictated by deeper theoretical reasons are less attractive, but nevertheless possible.
For instance, we include in Fig.~\ref{fig:plots} three $\kappa$ benchmarks, $\kappa =\pm1/2,3/2$, which could be obtained in such two-field extensions, hereafter denoted by
 $X_{\pm1/2,~3/2}$.

Since $\mathcal{O}_{lq}^{(1)}$ and $\mathcal{O}_{lq}^{(3)}$ affect the production of neutrinos and of charged leptons, the LEFT Wilson coefficients  will be likewise correlated.
Considering only the tree-level matching, the correspondence reads
\begin{equation}\label{eq:alphatau}
   \delta C^{\nu_\ell \nu_\ell}_L = \frac{\kappa_\ell-1}{\kappa_\ell+1} \delta C^{\ell \ell}_9 \,.
\end{equation}
We take the simplest assumption in the following, namely that \textit{$\kappa_e = \kappa_\mu = \kappa_\tau \equiv \kappa$ and that the operators are flavor diagonal in the lepton sector}. The first assumption does not impact significantly the conclusions since, as we will see, the dominant contribution to the processes stems from $\kappa_\tau$. On the other hand, the flavor diagonality in the lepton sector is an additional relevant requirement that we have to impose in order not to lose predictivity.
\end{section}
\begin{section}{Analysis and Discussion}
\label{sec:Analysis}
In the following we focus on the four processes $B^{0,+}\to K^{(0\ast),+} \tau^+ \tau^-$, $B^{0,+}\to K^{(0\ast),+} \nu \bar\nu$. Henceforth we will drop the electric charges. We will study in detail whether or not it is possible to conciliate the charged and neutral currents anomalies in light of the current experimental bounds on $B \to K^{(\ast)} \nu\bar{\nu}$ and  $B \to K^{(\ast)} \tau^+ \tau^-$~\cite{Belle:2017oht}, and the recent LHCb measurements of $R_{D^{(\ast)}}$~\cite{LHcb-partial}. We employ \textsc{Flavio}~\cite{Straub:2018kue} to express the branching ratios, reported in App.~\ref{App:BR}, in terms of the relevant Wilson coefficients listed in Eqs.~(\ref{eq:bsll-smeft})-(\ref{eq:left-CL-bsnunu}).
The branching ratios are computed in  the full kinematic range of $q^2$.
We stress that we will ignore running effects as we have checked that they are subdominant for the operators relevant to the processes of interest.

The deviation from the standard model of $R_{D^{(\ast)}}$ can be parameterised
by $\delta C_V^{\tau \nu}-\delta C_A^{\tau \nu}$~\cite{Capdevila:2017iqn, Capdevila:2017bsm, Iguro:2018vqb,Gomez:2019xfw}
\begin{equation}
\begin{split}
    \delta C_V^{\tau \nu}-\delta C_{A}^{\tau \nu}& = \sqrt{\frac{R_{X}}{R^{\rm SM}_{X}}}-1 =
    \begin{cases}
     0.096(43) & X=D\\
     0.059(26) & X=D^*
    \end{cases}\,.
    \end{split}
\end{equation}
Such values give the average
\begin{equation}
   (\delta C_V^{\tau \nu}-\delta C_{A}^{\tau \nu})_\text{avg.}= 0.069 (22) \,,
\end{equation}
which we will use as a reference value. Eq.~(\ref{eq:bslnu-smeft})  implies that, ignoring the running effects,
\begin{equation}
   \frac{ \big{[}C_{lq}^{(3)}\big{]}_{3323}}{\Lambda^2}=-0.045 (15)\,\text{TeV}^{-2}\,.
\end{equation}
Assuming a Wilson coefficient of $\mathcal{O}(1)$, this suggests a scale of $\Lambda \approx 4.7$ TeV.
Fixing $\big{[}C_{lq}^{(1)}\big{]}_{3323}= \kappa \big{[}C_{lq}^{(3)}\big{]}_{3323}$, from Eq.~(\ref{eq:bsll-smeft}) we get
\begin{equation}
    \delta C_9^{\tau \tau}=- \delta C_{10}^{\tau \tau}=-(\kappa +1)(28\pm 9)\,,
\end{equation}
\noindent and, from Eq.~(\ref{eq:alphatau}),
\begin{equation}
    \delta C_L^{\nu_\tau \nu_\tau}
    =-(\kappa-1)(28\pm 9)\,.
\end{equation}
The value of $ \delta C_L^{\nu_\mu \nu_\mu}$ can be extracted analogously from the global fits on $b\to s \mu \mu$.
Since $\delta C_9^{\tau\tau}$ generates a sizable Universal contribution to $b \to s \ell \ell$ at one-loop~\cite{Capdevila:2017iqn, Crivellin:2018yvo, Cornella:2021sby, Alguero:2022wkd}, we will consider the fit results where this Universal contribution is allowed. The latest fit provides $\delta C_{9}^{\mu\mu} = -0.36(7)$\cite{Alguero:2021anc}.
Moreover, at the scale $\mu=m_W$ we obtain the Universal coefficient to be $\delta C^U_9\approx -0.20(6)\times(\kappa_\tau+1)\left[1+1/4\ln{\left(\Lambda/4.7 \ \text{TeV}\right)}\right]\,$, which is in good agreement with the global fit on this scenario~\cite{Alguero:2021anc,Cornella:2021sby}. From these values, using Eq.~(\ref{eq:alphatau}), it is evident that the contribution from tau-neutrinos to $B \to K^{(\ast)} \nu\bar{\nu}$ is $\mathcal{O}(10^2)$ times larger than the contributions of the other two neutrino species.

In Fig.~\ref{fig:plots} we show $R_{\nu\nu}^{K^{(\ast)}}$ versus $R_{\tau\tau}^{K^{(\ast)}}$ for the different models listed in Table~\ref{tab:models}.
\footnote{It is important to mention that in principle for each model there are two branches of solutions deriving from inverting Eq.s~\ref{eq:RKtau} and \ref{eq:RKstartau} to write $\delta C_9^{\tau\tau}\left(R_{\tau\tau}^{K^{(*)}}\right)$. In the figure, we show only the branches with the sign that can in principle explain the $R_{D^{(*)}}$ anomalies.}
\begin{figure*}[t!] 
\centering%
\subfigure[{}\label{fig:npsp1nqnn}]%
{\includegraphics[width=0.42\textwidth]{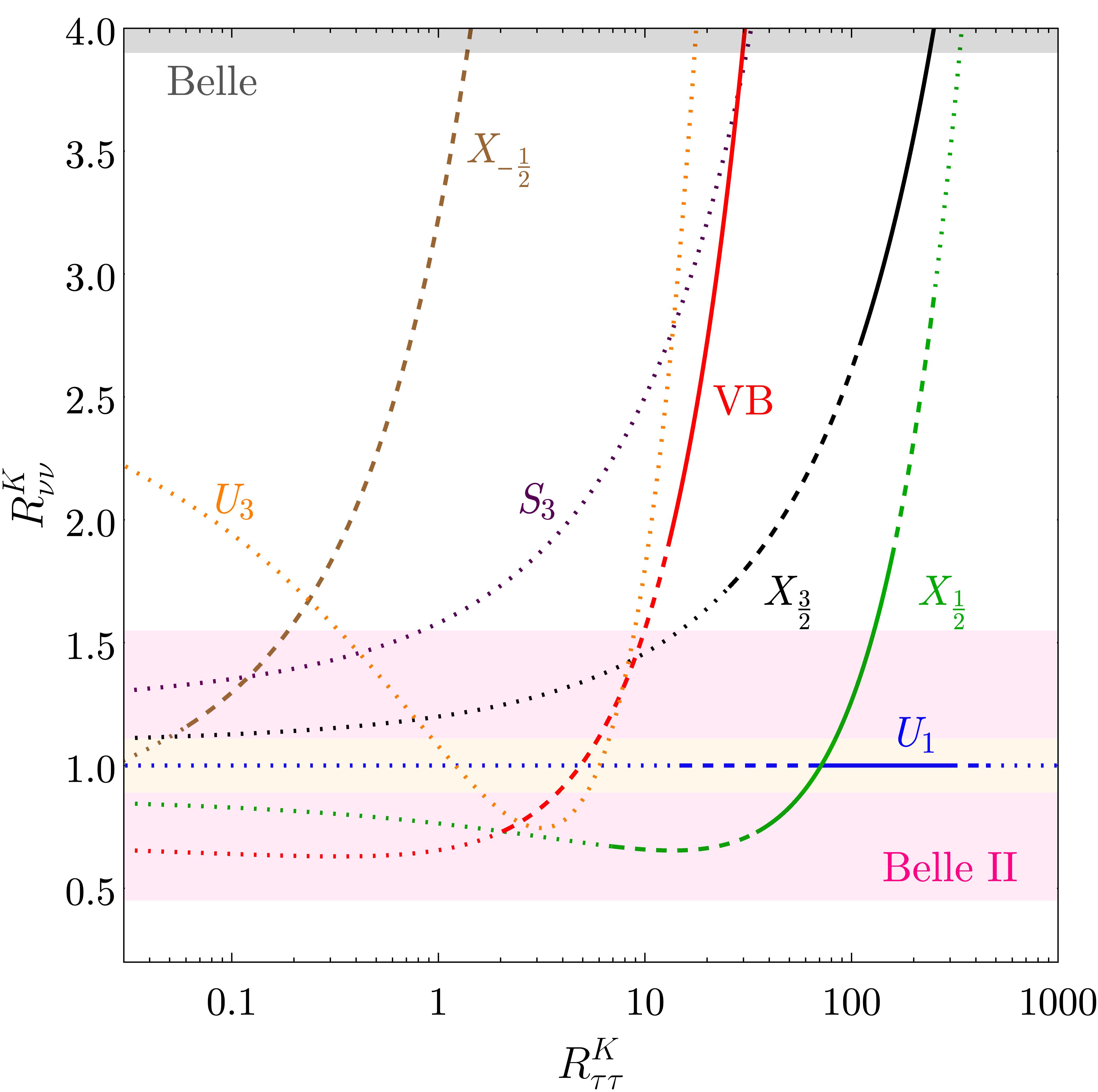}}\hspace{1cm}
\subfigure[{}\label{fig:npsp2nqnn}]%
{\includegraphics[width=0.42\textwidth]{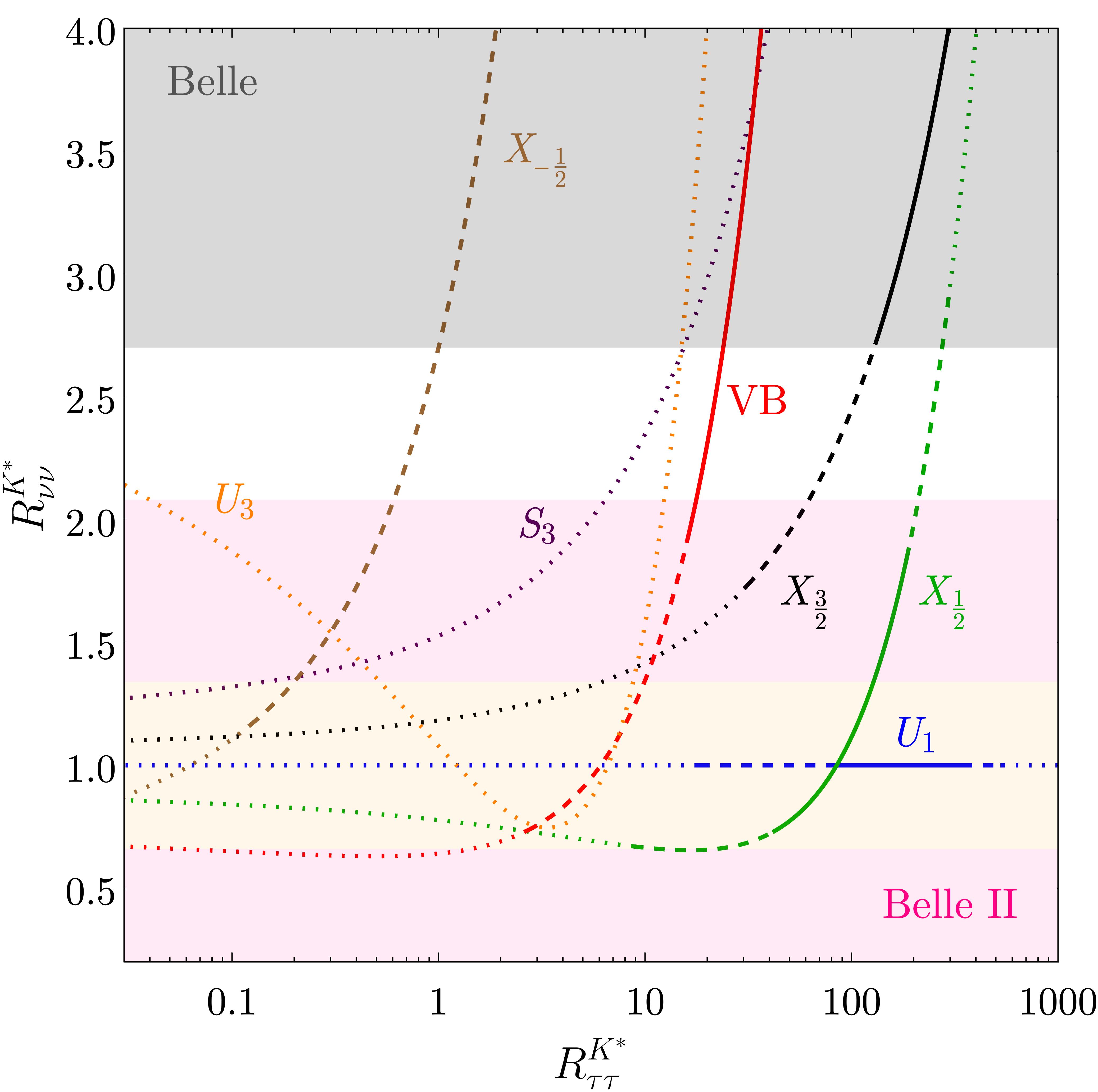}}
\caption{\small \sl  $R_{\nu\nu}^{K^{(\ast)}}$ as a function of $R_{\tau\tau}^{K^{(\ast)}}$ for different models. The solid (dashed) lines show the region where $R_{D^{(*)}}$ can be explained at $1\, (2) \sigma$, while the gray bands is the 90\% exclusion limit $B\to K^{(*)}\bar{\nu}\nu$~\cite{Belle:2017oht}. The baseline expectation of Belle II~\cite{Belle-II:2018jsg} at $1\ \text{ab}^{-1}\left(50\ \text{ab}^{-1}\right)$ is shown as pink(orange) shaded area.
\label{fig:plots}
}
\end{figure*}
As it can be seen, the two plots involving $K$ and $K^*$ are almost identical, due to the fact that we are neglecting right-handed couplings.

The fact that $R_{D^{(\ast)}}$ triggers a large contribution to $R_{\tau \tau}^{K^{(\ast)}}$ is a well known fact in the literature~\cite{Capdevila:2017iqn, Cornella:2021sby}.
The same source of NP affects also $R_{\nu\nu}^{K^{(\ast)}}$ unless there is a cancellation between $\big{[}C_{lq}^{(1)}\big{]}_{ij23}$ and $\big{[}C_{lq}^{(3)}\big{]}_{ij23}$ in Eq.~(\ref{eq:left-CL-bsnunu}). This is the case with the $U_1$ leptoquark, which predicts $R_{\nu\nu}^{K^{(\ast)}}=1$  (for a recent analysis see \cite{Aebischer:2022oqe}).
Nevertheless, the $U_1$ model gives a sizeable contribution to the production of taus such that at $95\%$ C.L.  
\begin{align}
    &15 \lesssim R_{\tau\tau}^{K^{(\ast)}} \lesssim 540\,.
\end{align}

The other leptoquarks considered in this work, $S_3$ and $U_3$, are already excluded by the bounds on dineutrino productions \cite{Angelescu:2018tyl}.
Interestingly, $U_1$ is not the only model that can evade the bounds on $R_{\nu\nu}^{K^{(\ast)}}$ while explaining $R_{D^{(\ast)}}$. 
 In fact, the VB can also achieve $R_{\tau\tau}^{(\ast)}\sim\mathcal{O}(10)$ with $R_{\nu\nu}^{K^{(\ast)}}\sim \mathcal{O}(1)$.
 However, this scenario is almost ruled out by other constraints involving $B^0_s-\bar{B}^0_s$ mixing and $\tau\to3\mu$ \cite{Bhattacharya:2016mcc}.

If we also consider two-field scenarios, we see that viable models generically have $ -1/2 \lesssim \kappa \lesssim 3/2$. 
In such a range, not only the contribution to the production of neutrinos can be kept reasonably small, but also the value of $R_{\tau\tau}^{K^{(*)}}$ can become as low as $\mathcal{O}(0.1)$.
If a signal was found to be compatible with the SM prediction, within the assumptions of this work the allowed range for $\kappa$ would be greatly reduced, either pointing to the $U_1$-model or to the requirement of having more than a single new field.
 
 The relation between $R^K_{\tau\tau}$ and $R^{K^\ast}_{\tau\tau}$ is also interesting. As can be seen from the expressions in the App.~\ref{App:BR}, considering only left-handed operators, i.e. $\delta C_9^{\tau\tau}=-\delta C_{10}^{\tau\tau}$, causes the variation in the two processes to deviate by the same order of magnitude and in the same direction. This is no longer true if the right-handed operators are included \cite{Buras:2014fpa}. 
 Indeed, if Belle II were to measure a large deviation in a process concerning $K$ but not in $K^\ast$, or vice versa, the simplest conclusion would be the presence of right-handed operators. Such analysis is, however, beyond the scope of this letter. 
\end{section}
\begin{section}{Conclusions}
\label{sec:Conlusions}
In this letter we have reviewed the possibility of a simultaneous explanation of $R_{D^{(\ast)}}$ and $R_{K^{(\ast)}}$, and the implication of such a scenario on the production of neutrinos and taus in $B\to K^{(*)}$ decays. In doing so, we have considered the most recent $R_{D^{(\ast)}}$ determination of LHCb~\cite{LHcb-partial} and the projected sensitivities of Belle~II.

In particular, we have considered different single-field BSM extensions  under the hypothesis of NP coupling only to Left-Handed fermions (see Sec.~\ref{sec:intro} for more details), and we have shown quantitatively their impact on $ R^{K^{(\ast)}}_{\nu\nu} $ and $ R^{K^{(\ast)}}_{\tau\tau} $. The results can be found in Fig.~\ref{fig:plots}.
If Belle II found a SM-like result for di-neutrinos production, then the preferred single-field extension would be the well-known $U_1$-leptoquark, with a consequent enhancement of $\mathcal{B}(B\to K^{(*)}\tau^+ \tau^-)$ of $\mathcal{O}(10^{2\divisionsymbol 3})$ with respect to the SM value. 
On the contrary, the space for models with at least two fields is far richer. In this regard, we have identified a range of possible models for which it is possible to obtain values in the $(R^{K^{(\ast)}}_{\tau\tau},R^{K^{(\ast)}}_{\nu\nu})$ plane inaccessible by single field extensions, e.g. even down to $\mathcal{O}(0.1)$ and $\mathcal{O}(0.6)$, respectively. 

Exciting experimental times are ahead of us, and the measurements of di-neutrinos and di-taus modes could be the game changer for the discovery of NP.
\end{section}
\section*{Acknowledgments} 
A.d.G. thanks Luca Merlo for useful discussions.
G.P. thanks Damir Be\v{c}irevi\'c and Olcyr Sumensari for  discussions and comments on this work.
This project has received funding /support from the European Union’s Horizon 2020 research and innovation programme under the Marie Skłodowska-Curie grant agreement No 860881-HIDDeN.
A.d.G. acknowledges as well support by the Spanish Research Agency (Agencia Estatal de Investigacion) through the grant IFT Centro de Excelencia Severo Ochoa No CEX2020-001007-S.
%
\newpage
\appendix
\numberwithin{equation}{section}
\begin{section}{Branching Ratios}
\label{App:BR}
In this section we list the relevant branching ratios normalised to the SM value as a function of the Wilson coefficients defined in Sec.~\ref{sec:treelevelmatching}. The full $q^2$ range is considered. Furthermore, we omit sub-leading contributions which are found to be $\mathcal{O}(5\%)$ or smaller. The Wilson coefficients of the following expressions are evaluated at the scale $\mu=m_W$.
\paragraph{\underline{\bm{$R_{\tau\tau}^{K^{(\ast)}}$}}}The expressions of $R_{\tau\tau}^K$ and $R_{\tau\tau}^{K^{\ast}}$ defined in Sec.~\ref{sec:intro} are
\begin{align}
    & \label{eq:RKtau} R_{\tau\tau}^K \approx 1+1.48\cross 10^{-2} \left(\delta C_9^{\tau\tau}-\delta C_{10}^{\tau\tau}\right)^2 \\
    &\nonumber+2.42\cross 10^{-1} \left(\delta C_9^{\tau\tau}-\delta C_{10}^{\tau\tau}\right)\\
    &\nonumber -1.15\cross 10^{-2} \left(\delta C_9^{\tau\tau}-\delta C_{10}^{\tau\tau}\right) \left(\delta C_9^{\tau\tau}+\delta C_{10}^{\tau\tau}\right)\\
    &\nonumber+1.48\cross 10^{-2} \left(\delta C_9^{\tau\tau}+\delta C_{10}^{\tau\tau}\right)^2\\
    &\nonumber-9.93\cross 10^{-2} \left(\delta C_9^{\tau\tau}+\delta C_{10}^{\tau\tau}\right) \,,
\end{align}
\begin{align}
    &\label{eq:RKstartau}  R_{\tau\tau}^{K^*} \approx 1+1.73\cross 10^{-2} \left(\delta C_9^{\tau\tau}-\delta C_{10}^{\tau\tau}\right)^2\\
    &\nonumber+2.61\cross 10^{-1} \left(\delta C_9^{\tau\tau}-\delta C_{10}^{\tau\tau}\right)\\
    &\nonumber+1.86\cross 10^{-2} \left(\delta C_9^{\tau\tau}-\delta C_{10}^{\tau\tau}\right) \left(\delta C_9^{\tau\tau}+\delta C_{10}^{\tau\tau}\right)\\
    &\nonumber+1.73\cross 10^{-2} \left(\delta C_9^{\tau\tau}+\delta C_{10}^{\tau\tau}\right)^2\\
    &\nonumber+1.26\cross 10^{-1} \left(\delta C_9^{\tau\tau}+\delta C_{10}^{\tau\tau}\right)\,.
\end{align}
\paragraph{\underline{\bm{$R_{\nu\nu}^{K^{(\ast)}}$}}} For $R_{\nu\nu}^{K^{(\ast)}}$,  neglecting $\delta C_L^{\nu_e \nu_e}$ we find
\begin{equation}
\label{eq:RKnu}
\begin{split}
        R^{K^{(*)}}_{\nu\nu}\approx&\,\, 1-  10^{-1} (\delta C_L^{\nu_\mu \nu_\mu}+\delta C_L^{\nu_\tau \nu_\tau})\\
        &+ 8.1\cross 10^{-3} \left[ \left( \delta C_L^{\nu_\mu \nu_\mu}\right){}^2+\left(\delta C_L^{\nu_\tau \nu_\tau}\right){}^2\right]\,.
\end{split}
\end{equation}
\end{section}
\bibliographystyle{BiblioStyle.bst}
\bibliography{Bibliography_aas.bib}
\end{document}